\documentclass[prl,twocolumn]{revtex4}

\usepackage{graphicx}
\usepackage{dcolumn}
\usepackage{bm}

\begin{document}


\title{The superconducting phase diagram 
in a model
for tetragonal and cubic systems with strong antiferromagnetic
correlations}
\author{J.J. Deisz}
\affiliation{Department of Physics \\University of Northern Iowa\\
Cedar Falls, IA 50614}

\date{\today}

\begin{abstract}
We calculate the superconducting phase diagram  
as a function of temperature and $z$-axis anisotropy
in a model for tetragonal and cubic systems having strong
antiferromagnetic fluctuations.
The formal basis for our calculations is 
the fluctuation exchange approximation (FLEX)
applied to the single-band Hubbard model near half-filling.
For nearly cubic lattices, two superconducting phase
transitions are observed as a function of temperature
with the low-temperature state having the time-reversal
symmetry-breaking form, $d_{x^2 - y^2} \pm id_{3z^2 -r^2}$. 
With increasing tetragonal distortion the 
time-reversal-symmetry-breaking phase is suppressed giving
way to only $d_{x^2 - y^2}$ or $d_{3z^2 -r^2}$
single-component phases. 
Based on these results, we propose that CeIn$_3$ is a 
candidate for exhibiting a time-reversal symmetry-breaking
superconducting state. 

\end{abstract}

\maketitle

The discovery of unusual superconductivity in the heavy fermion systems
focused much early effort on electronic pairing mechanisms and the possibility
that the resulting superconducting states have lower symmetry than the
underlying crystalline lattice. The competition between superconductivity
and magnetic ordered states suggests that spin fluctuations are a prime
candidate to form the glue that binds electrons into Cooper pairs. 
Power law dependencies of thermodynamic properties and phase diagrams that
include multiple superconducting states, 
for example the $H-T$ phase diagram of UPt$_3$ and the $x-T$
phase diagram of U$_{1-x}$Th$_x$Be$_{13}$ \cite{sigrist:1991a},
point to the lower
symmetry of the superconducting order parameter.  As a consequence of 
having a lower
symmetry in comparison to that for the lattice, 
the order parameter of an unconventional superconductor
can have more degrees of freedom than exhibited in a conventional
superconductor leading to phase diagrams involving multiple superconducting
states, but
microscopic models are needed
to shed light on the connection between
the pairing interaction and 
the pairing states that are produced.

Recently, the interplay between crystal 
lattice symmetry and spin-fluctuation
induced pairing instabilities 
was explored by
Monthoux and Lonzarich 
\cite{monthoux:1999a} and
Arita, Kuroki and Aoki \cite{arita:1999a}. 
They find that spin fluctuation-induced pairing
into the unconventional $d_{x^2-y^2}$ state
is most effective for producing
large transition temperatures ($T_c$)
for quasi two-dimensional lattices, 
a result that is consistent with the observation that the highest
$T_c$ values occur in the quasi two-dimensional cuprates and the
more recent finding that 
$T_c$ in the cubic heavy fermion system CeIn$_3$ 
\cite{grosche:1996a} is about one
order of magnitude less than is observed in a collection of
related quasi two-dimensional compounds, such as the series 
Ce$_n$T$_m$In$_{3n+2m}$
where T = Rh or Ir and $n$ = 1 or 2 and $m = 1$
\cite{thompson:2001a}.

In this Letter we report on model
calculations of the
superconducting phase diagram
as a function of temperature and tetragonal anisotropy
for electrons paired via spin fluctuations
in nearly antiferromagnetic systems.
The phase diagram is obtained numerically through 
the proper generalization of the
fluctuation exchange approximation for
the single band Hubbard model to the superconducting
state, a generalization which is necessary to
resolve the relative stability of
nearly degenerate unconventional pairing states.
We find that for cubic lattices
the stable superconducting state has the
time-reversal symmetry breaking
form $d_{x^2-y^2} \pm id_{3z^2-r^2}$. 
Small tetragonal distortions lift the degeneracy
between $d_{x^2-y^2}$ and $d_{3z^2-r^2}$ pairing states
leading to 
two superconducting phase transitions as a function
of decreasing temperature, but as the degeneracy is lifted further
through
larger tetragonal distortions the second transition is suppressed.
On the basis of these results, we suggest that the low-temperature
superconducting state in simple cubic
systems with strong antiferromagnetic correlations,
such as CeIn$_3$, are strong candidates for realizing
multiple superconducting phases including
a low-temperature phase with broken time-reversal symmetry.

The microscopic basis for our calculations is the
single-band Hubbard
model,
\begin{equation}
\label{hubbard}
H = -\sum_{i,j,\sigma}\left(t_{ij}\,
                      c^{\dagger}_{i,\sigma}c_{j,\sigma} + \mathrm{h.c.}
\right)
+ U \sum_i c^{\dagger}_{i,\uparrow}c_{i,\uparrow}
        c^{\dagger}_{i,\downarrow}c_{i,\downarrow},
\end{equation}
where $t_{ij}$ represents the electron hopping amplitude
between sites $i$ and $j$ and $U$ is the on-site 
interaction energy between up and down spin electrons.
The values of $t_{ij}$ reflect the underlying lattice
structure.
For tetragonal lattices, the hopping parameter for
unit displacements in the $x$ or $y$ directions, $t_{xy}$, 
is distinct from the same for unit displacements
along the $z$ axis, $t_z$.  For simplicity we set
the hopping integrals equal to zero for larger displacements 
in which case the non-interacting electron bandwidth is equal to
$W_0 = 8 t_{xy} + 4 t_z$.  

An approximation scheme must be employed for 
calculations based on this model.
We use the fluctuation exchange approximation (FLEX) of
Bickers, White, and Scalapino \cite{bickers:1989a},
a numerically-based scheme
that is conserving in the sense described by Baym \cite{baym:1962a}.
FLEX provides a self-consistent description of both the 
quasiparticles and the magnetic-fluctuation
induced pairing interaction for a given
on-site interaction strength, temperature, and band filling.
It is expected that
results obtained with FLEX are 
quantitatively accurate for, at best, 
weak-to-intermediate coupling, \textit{i.e.} $U / W_0 \alt 1$
\cite{bickers:1991a, freericks:1994a}.  However,
even within this range
there are notable qualitative failures such as, as
will be discussed shortly,
violations of the Mermin-Wagner-Hohenberg theorem
for $d$-wave superconducting states.

In contrast to other works, our formulation of FLEX includes the entire
set of fluctuation diagrams for the electron self-energy
in both the normal and superconducting states,
\textit{i.e.} we include all possible combinations of particle-like, hole-like
and anomalous Green's functions.
At the formal level, this ensures that response functions
and Green's functions are obtained in a consistent manner 
such that conservation laws derivable from symmetries of the Hamiltonian are
obeyed.  While $T_c$ is determined by
self-energy diagrams which contain
only one anomalous Green's function,
a more complex set of diagrams with three anomalous Green's functions
contribute to fourth order
terms in the Ginzburg-Landau expansion for the free energy 
in terms of the superconducting order parameter and
these terms are known to
determine the relative stability of multicomponent pairing
states \cite{sigrist:1991a}.  
We note that the results we obtain for 
$T_c$ in two dimensions are in agreement
at the 10\% level with those obtained earlier \cite{pao:1994a}
suggesting that self-energy diagrams that are 
omitted in most treatments of FLEX do not
play a large role in determining $T_c$,
at least in the two-dimensional limit.

\begin{figure}
\includegraphics{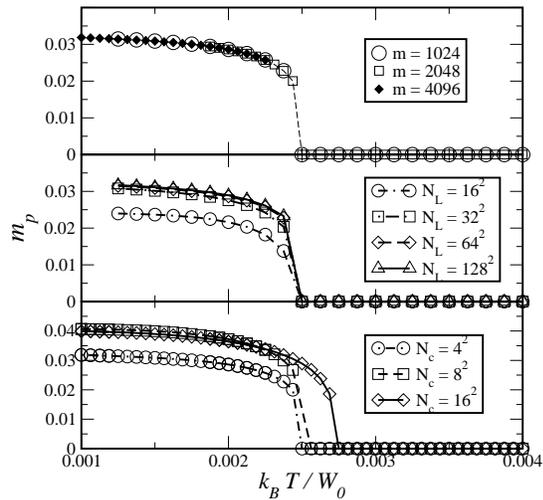}
\caption{\label{fig:systematics}
Numerical systematics in the evaluation of
pairing amplitude, $m_p$, versus the scaled temperature, $k_B T / W_0$,
where $W_0$ is the non-interacting electron bandwidth, for
two-dimensional lattices with  
0.85 electrons/site and $U / W_0 = 0.5$.
Top: Variation as a function
of number of Matsubara frequencies, $m$. 
Middle: Dependence on
the number of lattice points, $N_L$.
Bottom graph:  Variation as a function
of DCA cluster size, $N_c$.}
\end{figure}

To access sufficiently large system sizes with modest computational resources,
we combine this formulation of FLEX with the dynamical cluster approximation
(DCA) \cite{hettler:2000a}.
Essentially, in the DCA large lattice calculations are
made feasible by
approximating correlation effects via a smaller embedded cluster.
When combined with the DCA, there are three numerical
parameters in FLEX: the number of Matsubara frequency points
used ($m$), the lattice size ($N_L$) and the DCA cluster
size ($N_c$). 
We show the dependence of results for the pairing amplitude,
$m_p$ (defined below), versus temperature on these
numerical parameters
in Figure~\ref{fig:systematics} 
for a two-dimensional lattice.
For the range of these parameters that we can access,
the most significant variation in these curves
is through the DCA cluster size (bottom graph 
in Figure~\ref{fig:systematics}).  
Nonetheless, we find  
that the error made in the result for $T_c$ is only on the
order 10\% when using a $4^2$ DCA cluster.
In what follows, we use $m=16384$ frequency points,
$N_L = 32^3$ lattices and 
$N_c = 4^3$ DCA clusters.  

Although a variety of electronic and thermodynamic properties
can be calculated with the FLEX,
our main focus here is on results for
the superconducting
transition temperature and the associated order parameter symmetry.
We obtain these results by determining the spatial dependence
of the pair wave function, $\psi(\mathbf{r})$,
for the stable
superconducting state and
the associated pairing amplitude, $m_p$.
These quantities are obtained
from the self-consistent result for
the anomalous Green's function, $F$, via
\begin{equation}
F_{\downarrow\uparrow}(\tau \to 0^-, \mathbf{r}) \equiv 
\langle c_{\mathbf{r}=\mathbf{0},\uparrow} c_{\mathbf{r},\downarrow}
\rangle = 
m_p \, \psi(\mathbf{r}).
\end{equation}
Other than requiring that $\psi({\mathbf{r}})$ is
even as a function of $\mathbf{r}$ 
(\textit{i.e.} we restrict ourselves
to singlet-pairing), no assumption is made
on the symmetry of the pairing state.
To permit any possible singlet-pairing state to emerge
in these calculations,
we initialize the self-consistent FLEX equations with
a small, spatially random pairing field to induce
a pairing amplitude in all possible symmetry channels.
The small field is removed as the self-consistent procedure
projects the wavefunction of the most stable
pairing state, $\psi({\mathbf{r}})$. 

We emphasize the mean-field nature
of the FLEX phase diagram for $d$-wave superconductivity
in the single-band Hubbard model.  The
mean-field $T_c$ values essentially
represent the temperature at which 
superconducting order emerges locally.  The true thermodynamic
$T_c$ is determined by phase fluctuations that are not
present in FLEX.  
The neglect of these 
fluctuations is especially problematic
in the two-dimensional
limit where they are known to eliminate the possibility of a
finite temperature phase transition.
Nonetheless, it is well known that a relatively weak interplanar
coupling can stabilize superconductivity in a quasi-two-dimensional
systems and the mean-field phase diagram is revealing
with respect to determining the conditions under which
the tendency toward superconducting
order is greatest.

\begin{figure}
\includegraphics{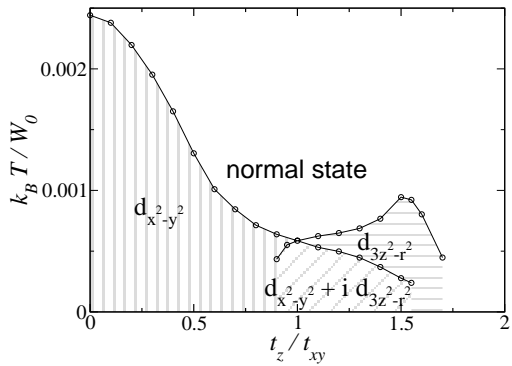}
\caption{\label{fig:phase} Fluctuation exchange approximation
results for the mean-field superconducting
phase diagram of the Hubbard model
as a function of the
ratio of the interplanar to intraplanar hopping,
$t_z / t_{xy}$, and reduced temperature, $k_B T / W_0$, 
for 0.85 electrons/site and $U / W_0 = 0.5$.
The $d_{x^2 - y^2}$ and $d_{3z^2 -r^2}$ pairing states are most
stable for
$t_z / t_{xy} < 1$ and $t_z / t_{xy} > 1$ respectively.
For $ t_z / t_{xy} \sim 1$ a second transition occurs
as a function of decreasing temperature into
the time-reversal symmetry breaking state
$d_{x^2 - y^2} + i d_{3z^2 -r^2}$.}
\end{figure}

Our primary result, shown in Figure~\ref{fig:phase}, is the
calculated superconducting phase diagram
as a function of the scaled temperature, 
$k_B T / W_0$, and
the ratio of interplanar to intraplanar
hopping, $t_z/ t_{xy}$ for fixed values of
density ($n = 0.85$ electrons per site) and the ratio
of the on-site interaction energy to bare electron bandwidth
($U / W_0 = 0.5$).
We focus on three features 
in Figure~\ref{fig:phase}:
the low-temperature stability of the time-reversal symmetry breaking
state $d_{x^2-y^2} \pm i d_{3z^2-r^2}$
for lattices with cubic and nearly cubic symmetry,
two distinct superconducting transitions as a
function of temperature for nearly cubic lattices,
and, in agreement with previous results,
maximal $T_c$ values occurring in the quasi-two-dimensional
limit.

We first focus on the cubic limit, $t_z / t_{xy} = 1$,
for which $T_c$ is at a local minimum.
We find that, in agreement with Arita, Kuroki and Aoki \cite{arita:1999a},
that the stable superconducting state belongs to
the two-fold degenerate representation $\Gamma_3^+$ of the cubic
group which is described by basis functions of the
form $d_{x^2-y^2}$ and $d_{3z^2-r^2}$.  Symmetry considerations
allow for either single-component or multicomponent states 
of the form
$d_{x^2-y^2}$, $d_{3z^2-r^2}$, $d_{x^2-y^2} \pm d_{3z^2-r^2}$,
or $d_{x^2-y^2} \pm i d_{3z^2-r^2}$.
Weak-coupling-based arguments suggest that the most stable state
is the one for which the superconducting gap is most complete
on the Fermi surface as this will tend to maximize the
condensation energy.  This argument favors the  
$d_{x^2-y^2} \pm i d_{3z^2-r^2}$ pairing state
as it has point nodes 
while the others have line nodes \cite{sigrist:2000a}.

FLEX incorporates feedback effects between
quasiparticles, the order parameter and the pairing
interaction and, thus, does not necessarily 
generate the pairing state expected from weak-coupling
theory.  Nonetheless, 
we indeed find that FLEX produces
the $d_{x^2-y^2} \pm i d_{3z^2-r^2}$ pairing state
in this case.  This state
breaks time reversal
symmetry leading to unusual phenomena such as
as bulk magnetic effects associated with
the superconducting pairs \cite{sigrist:1991a}.
Assuming that quasiparticles are well-defined in
the superconducting state, such pairing is expected
to generate thermodynamic properties that
reflect the point node structure of the gap function,
such as having a $T^3$ low-temperature
specific heat.

For slight deviations from cubic symmetry,
\textit{i.e.} $t_z / t_{xy} = 1 \pm \epsilon,\;
\epsilon \ll 1$, $T_c$ increases due to the
improved stability of either
the $d_{x^2-y^2}$ or $d_{3z^2-r^2}$ pairing state.
The two pairing states are no longer degenerate
for $\epsilon \neq 0$
and the initial transition from the superconducting
transition is into a single component state.  However,
because of 
the near degeneracy of these states, a
second superconducting transition occurs 
into the $d_{x^2-y^2} + i d_{3z^2-r^2}$ state with
decreasing temperature.  
The low-temperature state is nodeless 
because $d_{3z^2-r^2}$ basis functions
are members of the symmetric group for tetragonal lattices
and, consequently, are intrinsically
mixed with contributions from basis functions
with $s$-wave symmetry.
However, $s$-wave terms in the pair wavefunction are relatively
small.  For example,
the $s$-like terms have amplitudes which are about
4\% of those for $d_{3z^2-r^2}$
for $t_z / t_{xy} = 1.1$.
Consequently, the gap is correspondingly
small for a set of points on the Fermi surface
that correspond to the point nodes that are found
for cubic lattices.

As is expected, we find that for lattices
favoring in-plane conduction, \textit{i.e.} 
$0 \le t_z/ t_{xy} < 1$, the
$d_{x^2-y^2}$ pairing state is most stable, 
while the $d_{3z^2-r^2}$ (+ $s$-wave) pairing state
is most stable when interplanar motion is enhanced,
\textit{i.e.} $ t_z / t_{xy} > 1$.
It is apparent from Figure~\ref{fig:phase} that
the largest $T_c$ values are obtained in
the quasi-two-dimensional limit, \textit{i.e.} $t_z/ t_{xy} \to 0$,
in agreement with the results referred to earlier 
\cite{monthoux:1999a, arita:1999a} and in
line with the trend observed in the compounds
CeIn$_3$ \cite{grosche:1996a} and
Ce$_n$T$_m$In$_{3n+2m}$
where T = Rh or Ir and $n$ = 1 or 2 and $m = 1$ \cite{thompson:2001a}.
There also is a region of enhanced $T_c$ 
for $t_z / t_{xy} > 1$ up to approximately
$t_z / t_{xy} \sim 2$ at 
which point $T_c$ becomes vanishingly small.  We note that 
$t_z / t_{xy} = 2$ corresponds to the point at which
the interplanar bandwidth equals the bandwidth corresponding
to in-plane motion.

It is interesting to consider these results
in light of experimental data for superconductors
with cubic symmetry.
Data for the compound PrO$_4$Sb$_{12}$ is consistent
with the superconducting phases observed in this calculation
\cite{goryo:2003a}.  
For example, specific heat measurements show
evidence of a double superconducting transition
\cite{vollmer:2003a} and muon-spin relaxation data is consistent
with a low-temperature time-reversal symmetry breaking
state \cite{aoki:2003a}.  However, the strong collective mode
that forms the glue between paired electrons is due, most likely,
to electric quadrupolar rather than magnetic degrees of
freedom \cite{kohgi:2003a}.  Nonetheless, it is interesting
to note the possibility that two different pairing mechanisms
may tend to generate the same time-reversal symmetry breaking
state in cubic symmetry and,
thus, the phenomenology developed for the 
superconducting state in PrO$_4$Sb$_{12}$ \cite{goryo:2003a}
may apply also for magnetically-paired superconductors.

The alloy series U$_{1-x}$Th$_{x}$Be$_{13}$ displays a 
complex superconducting phase
diagram as a function of thorium concentration, $x$.
For  $0.2 \alt x \alt 0.4$ values,
a double superconducting transition is observed
in the electron specific heat \cite{ott:1985a} and
muon spin resonance data points to the appearance
of an internal magnetic field below the second transition.
Sigrist and Rice interpreted the magnetic anomaly
in terms of a multicomponent time-reversal symmetry breaking
superconducting state \cite{sigrist:1989a}.
However, Kromer \textit{et al.} \cite{kromer:1998a}
use specific heat and lattice
expansion data to argue for the appearance of spin
density wave below $T_c$.
The interpretation of the phase diagram for these
compounds is still being debated \cite{martisovits:2000a}.
We simply note that the scenario described by
Sigrist and Rice is consistent with pairing states
that are observed in the model 
that we consider here and should their interpretation
be incorrect it would point to the importance of processes
that are neglected in spin-fluctuation models.

The cubic superconductor
CeIn$_3$ is, perhaps, the most likely candidate
system for connecting with these model calculations.
On account of the vicinity of the Ne\'el
state, there is a strong likelihood that
electron pairing is related to the strong
antiferromagnetic correlations in this system.
To the best of our knowledge, no evidence has been presented
either for a double superconducting transition as
a function of temperature nor for a low-temperature
time reversal symmetry breaking state in this compound.
In light of weak coupling arguments \cite{sigrist:2000a}, 
the FLEX results presented here and the anomalies observed
in other unconventional cubic superconductors, the presence or
absence of time-reversal symmetry breaking in this
compound will be revealing with respect to the applicability of
simple models, such as the one considered here, for 
providing a minimal microscopic basis for understanding
superconductivity in CeIn$_3$ and related compounds.

In summary, we have performed fluctuation exchange
approximation calculations
for the Hubbard model near half-filling to model 
the phase diagram of tetragonal
and cubic superconducting systems whose pairing is mediated
by antiferromagnetic spin fluctuations. 
Near cubic symmetry,
two phase transitions are realized, the first corresponds
to a transition to a single-component state and the second
transition occurs to a multicomponent state with broken
time reversal symmetry.
On the basis of these calculations, we propose that
CeIn$_3$ is a candidate material for exhibiting
broken time-reversal symmetry in the superconducting state.

\begin{acknowledgments}
We are especially grateful to D.W. Hess for many useful conversations.
The author extends his appreciation to the Department of Physics
at the University of Cincinnati for their hospitality
during the summer of 2002 and to  
M. Jarrell and T. Meier for useful conversations
during that time.
The author gratefully acknowledges partial support from
the Graduate College of the University of Northern Iowa.
\end{acknowledgments}


\end{document}